\documentclass[preprint2]{aastex}

\usepackage{epsf}
\usepackage{graphics}
\usepackage{color}

\newcommand{\bl}{$\bullet$}
\newcommand{\be}{\begin{equation}}
\newcommand{\ee}{\end{equation}}

\newcommand{\rb}[1]{\raisebox{1.5ex}[-1.5ex]{#1}}
\newcommand{\msun}{$M_{\odot}$}

\shorttitle{Emission Line Multiplet Method}
\shortauthors{Grupe, Pradhan & Frank}

\begin{document}


\def\etal{{\it et\thinspace al.}\ }
\def\alp{{$\alpha$}\ }
\def\al2{{$\alpha^2$}\ }

%
%
%


\title{Studying the Variation of the Fine Structure Constant Using Emission Line
Multiplets
\thanks{Based on observations obtained at MDM Observatory, Arizona}
}


\author{Dirk Grupe\altaffilmark{2}, 
Anil K. Pradhan \,and Stephan Frank}
\affil{Department of Astronomy, The Ohio State University,
    140 W. 18th Ave., Columbus, OH-43210, U.S.A.}
\email{dgrupe, pradhan, frank@astronomy.ohio-state.edu}
\altaffiltext{2}{Current address: Astronomy Department, Pennsylvania State
University, 525 Davey Lab, University Park, PA 16802; email:
grupe@astro.psu.edu}




\begin{abstract}
 As an extension of the method by Bahcall et al. (2004) to investigate the
time dependence of the fine structure constant, we
describe an approach based on new observations of forbidden line multiplets
from different ionic species. We obtain optical spectra of fine
structure transitions in [Ne~III], [Ne~V], [O~III], [OI], and [SII]
 multiplets from
a sample of 14 Seyfert 1.5 galaxies in the low-z range 0.035 $< z < 0.281$.
Each source and each multiplet is independently analyzed to ascertain
possible errors. Averaging over our sample, we obtain a conservative value
 $\alpha^2(t)/\alpha^2(0)$ = 1.0030$\pm$0.0014. However, our sample is 
 limited in size and our fitting technique simplistic as we primarily intend 
 to illustrate the
scope and strengths of emission line studies of the time variation 
of the fine structure
constant. The approach can be further extended and generalized to a
"many-multiplet emission line method" analogous in principle to the
corresponding method using absorption lines. With that aim,
we note that the theoretical limits on emission line ratios of selected ions
are precisely known, and provide well constrained selection
criteria. We also discuss several other forbidden
and allowed lines that may constitute the basis for a more rigorous study
using high-resolution instruments on the next generation of 8 m class
telescopes.
\end{abstract}

\keywords{galaxies: active - Atomic data
}

\section{Introduction}

 The variation of the fine structure constant \alp=$e^2/4\pi \epsilon_0 h c$
 is of fundamental interest in cosmology. 
  However, if there is a variation of the
 fundamental constants by time, the effect will be very small. Recent laboratory
 measurements using $\rm ^{171}Yb^+$ by \citet{peik04} give an upper limit of
 2$\times 10^{-15} \rm yr^{-1}$ resulting in a change of \alp~ of the order of
 $10^{-5}$ in 10 Gyrs. In order to measure this effect using astronomical
 observations, they have to be performed with high-redshift objects. Therefore
 the lines to be studied need to manifest themselves as strong features in the spectrum.
 The most common methods are measurements of high-redshift Ly$\alpha$ forest
  metal absorption lines.
Among the first ones was the ``alkali-doublet method"
\citep[e.g.][]{bahcall67}
with transitions from the singlet ground level to the two fine
structure doublets in alkali-like systems, such as the recent study of
C~IV and Si~IV systems using a QSO spectra from UVES
\citep{fiorenzano03}. The most extensive body of
work in recent years is described by
\citet{murphy03} (and references therein),
who have considerably extended absorption line studies to
a relatively large number of multiplets in heavy atomic systems,
including the iron group elements. Their ``many multiplet method" represents
`an order of magnitude increase in precision' over the alkali doublet
method and uses 
Lyman$\alpha$ forests of Quasars \citep[e.g.]{dzuba99a, dzuba99b, webb99,
webb02, murphy01a, murphy01b}.

 Recently, as an alternative 
 \citet{bahcall04} described an extensive analysis of
forbidden fine structure lines of the well known nebular [O~III] doublet
at 5006.84 and 4958.91 \AA.
These lines have the great advantage that they are extremely bright and
nearly ubiquitous in the optical spectra of H~II regions in most
sources. The approach by \citet{bahcall04} involved the study of archived
spectra of 3,814 quasars from the large database of the Sloan Digital Sky
Survey \citep[SDSS][]{york00}. They devised
elaborate algorithms based on stringent selection criteria to search for
and obtain a standard spectral sample of 42 quasars, as well as
alternate samples based on variants of the selection criteria.
Among these criteria are fits to the line profiles and line ratios.
Since both lines originate within the same upper level, their line
profiles must be similar and their line ratios depend only on intrinsic
atomic properties, the energy differences between the fine structure
levels and corresponding spontaneous decay Einstein A-coefficients, i.e.
independent of external physical conditions such as the density,
temperature, and velocities. \citet{bahcall04} also noted that other similar
forbidden multiplets of [Ne~V] and [Ne~III]
 could be exploited for emission line studies, but they are likely to be
weaker than the [O~III] by an order an magnitude in typical quasar
spectra.

The quasar absorption line many-multiplet method has clear advantages:
 because it measures
the Ly$\alpha$ forest of high-redshift systems at observed optical wavelengths
its look-back time into the Universe's history is long.  It also uses
many line pairs to measure any time dependence of
$\alpha$ and provides good statistics to minimize systematic errors of
the measurements of an individual line pair.
However, the method has some disadvantages: It is observationally 
expensive, requiring very high
resolution, like $\lambda/\Delta\lambda \geq$50000 using high-resolution
Echelle spectrographs and long exposure times on large telescopes 
(e.g. \citet{murphy03}).
Also the line pairs measured by this
method are from different atomic stages suggesting that they are not 
necessarily formed
in the same regions. Another disadvantage is that the large number of
absorption lines in Ly$\alpha$ forest systems make it difficult for the
correct identification of the lines; many lines are blended requiring
multi-Gaussian fits to the line blends in order to determine the wavelengths.
Based on high-resolution Keck spectra of several different
samples of quasars, \citet{murphy03} report a 'highly significant' value of
$\Delta \alpha/\alpha_0$ = $-0.57 \pm 0.10 \times 10^{-5}$. In contrast
the statistically invariant result by \citet{bahcall04} yields a value 
of $0.7 \pm 1.4
\times 10^{-4}$. An extensive discussion of relative problems and advantages
of various methods is given, among others, by \citet{bahcall04}
and \citet{levshakov03}.

 The optical [O~III] forbidden
emission lines $\lambda\lambda$4959,
5007\AA~ studied by \citet{bahcall04} are relatively easy to measure in
most AGN,
(except in Narrow-Line Seyfert 1 galaxies (NLS1s) in which these lines can be
very weak and contaminated by strong FeII emission 
\citep[e.g.][]{bor92,gru04,sp03,spn04}.
The observations can be performed with lower resolution Spectrographs and
smaller telescopes. \citet{bahcall04} presented a sample of 44 AGN carefully
selected from the SDSS and measured the wavelength
shifts between the [OIII]$\lambda\lambda$4959,5007\AA~ lines.
The disadvantage of this method is that it only uses one line pair and it can
only be used in the optical wavelength range for objects with redshifts z$<$0.8
before the [OIII]$\lambda$5007\AA~line is shifted out of the observable optical
wavelength range and NIR spectroscopy is required for objects of higher
redshift.

 In this paper we generalize the emission line method
outlined by \citet{bahcall04} by incorporating, in principle, the main
advantage of the many-multiplet absorption line method.
However, there are other key differences with \citet{bahcall04}.
We carry out new observations of quasars and Seyfert 1 galaxies, rather
than analyze existing data. This obviates the need for complex search
routines that forms the bulk of the analysis of the SDSS data by
\citet{bahcall04}. We expect better accuracy since 
 objects are pre-selected from previous observations for good quality
optical spectra to enable optimum analysis. Since the redshifts are
known a priori, we know the approximate positions of the well known forbidden
lines. Theoretically known line ratios (discussed in the next section)
offer an additional indication of the quality of the spectra.

 This many-multiplet emission line  method not only allows us to
measure more than one line pair per object and gives a more secure result, it
also, once established, opens the possibility
of observing objects of higher redshift in the optical wavelength range. Another
purpose of our observing run was to test whether it is feasible to use a 2m
class telescope with a medium resolution spectrograph to get enough accuracy to
perform a measurement of the time variation of the
fine-structure constant $\alpha$. While the
availability of observing time at large 8-10m class telescope is very limited,
many institutions have access to 2-3m small/medium size telescopes.
In this work we focus
on laying out the framework for general emission line analysis, potentially
leading to future studies using the many-multiplet emission line method, 
with a relatively
small sample. However, we discuss a variety of further extensions.

This paper is organized as follows: in \S\,\ref{theory} we describe the
theoretical background of the fine structure separation of the LS multiplets,
in \S\,\ref{observe} we describe the
spectroscopic observations at MDM observatory, in \S\,\ref{results}
the results
are presented, and discussed in \S\,\ref{discuss} usually in the context of
systematical errors in the wavelength calibration. 
Finally, we note a few main features of our method in the concluding 
section\,\ref{conclude}.

\section{\label{theory} Theory}

Relativistic electron
interactions lead to fine structure separation of LS-coupling
multiplets.
The level energies may be expanded in terms of a non-relativistic term
and relativistic terms in powers of \alp \citep{bs77}.
Although the exact formulation is predicated on the Dirac theory
\citep{grant96}, Breit formulated the generalized interaction for
non-hydrogenic systems that leads to
the so called Breit-Pauli approximation for the total electronic
Hamiltonian as a sum of one-body and two-body operators, referred to as
the non-fine structure and fine structure terms \citep{drake96, ff96}. 
The leading relativistic terms, such as the spin-orbit
coupling, are of the order of \alp$^2$; higher order terms
are orders of magnitude smaller. Forbidden lines arise from transitions
between LS states of the
same electronic configuration of an atomic system, usually the ground
configuration. Fine structure components of a forbidden LS multiplet are
separated by $\Delta$E $\propto$ \alp$^2$. The time dependence of \alp
therefore should manifest itself in different energy or wavelength
separations at different cosmological epochs.

\subsection{Forbidden Lines and \alp(t)}

 \citet{bahcall04} show that, for two lines from a forbidden
multiplet, the ratio

\be
     R(t) = \frac{\lambda_2 - \lambda_1}{\lambda_2 +
\lambda_1},
\ee
at a cosmological time t is related to the ratio at the present epoch t
= 0 by
\be
     \frac{R(t)}{R(0)}  = \frac{\alpha^2(t)}{\alpha^2(0)},
\ee

where the RHS is a measure of the variation in \alp as a function of
the 'cosmological look-back time' t. Wavelength separation between a
pair of well defined forbidden lines may thus be used as a
`chronometer' {\it provided}
its value at an earlier cosmological epoch differs from that at
present.

As mentioned, \citet{bahcall04} had already suggested the possibility of
extending this study with
other emission line pairs, such as [Ne~V]$\lambda\lambda$3346,3426\AA~ and
[Ne~III]$\lambda\lambda$3869,3968\AA.  The left panel of
Figure\,\ref{linescheme}
shows the schematic diagram
of four ionic species ([NeIII], [NeV], [OIII], and [OI])
and the two lines $^1D_2 \rightarrow ^3P_2$ and
$^1D_2 \rightarrow ^3P_1$. It might be noted
that while [O~I] and the isoelectronic ions C-like [O~III] and [Ne~V] have the
ground level $2p2 (^3P_0)$, the O-like [Ne~III] has a different atomic
structure with the ground level $2p4 (^3P_2)$. In addition we consider the
well known nebular doublet lines [OII] and [SII] due to transitions 
$^2D^o_{5/2},^2D^o_{3/2} \longrightarrow ^4S^o_{3/2}$.
The atomic transitions within the ground configuration correspond to
magnetic dipole (M1) and the much smaller electric quadrapole (E2)
interactions. The wavelengths and atomic transitions of the [NeIII], [NeV],
[OIII], [OI], and [SII] lines are given in Table\,\ref{line_wave}.

\subsection{Line Ratios}

 Subsequent analysis of different emission line multiplets rests on the
basic property of the ratio of lines originating in the same upper
level. For such a 3-level system the theoretical emissivity ratio of
line intensities is

\be
     LR = \frac{N_3 A_{31} h\nu_{31}}{N_3 A_{32} h\nu_{32}}.
\ee

 The upper level population is the same for both lines, and the $
R_{\epsilon}$ depends only the intrinsic atomic parameters, the
transition rates or A-values, and the energy differences. Generally, for
forbidden lines, the A-values are obtained from sophisticated theoretical
calculations, whereas the energies are available from laboratory
measurements. While the prime resource for both quantities is the on-line
National Institute of Standards
and Technology (NIST) compilation (www.nist.gov), some of the NIST data
for A-values is not quite up-to-date. The differences may be slight but
quite significant. Table\,\ref{line_ratio} 
shows the line ratios for the 5 ions and
respective multiplets using both the NIST data and the most recent
calculations.

 In their work \citet{bahcall04} quote the line ratio for the
[O~III] $\lambda\lambda$4959,5007\AA~ line using NIST values as 2.92, as
opposed to their measured value of 2.99 $\pm$ 0.02. However, the 2.92 is
the ratio of the A-values alone, not the emissivity ratio above; taking
account of the energy differences the NIST tabulations yield 2.89. While
this is in even worse agreement with the measured ratio, the most recent
calculations of A-values by \citet{sz00} yields a
significantly better theoretical value of 2.98, in agreement with
\citet{bahcall04}. The crucial difference between the \citet{sz00}
results and previous calculations \citep[such as by][]{galavis97} is
taking account of higher order relativistic corrections to the magnetic
dipole M1 operator \citep{drake71, ez81}. The above
discussion emphasizes the need for extremely accurate atomic
calculations including all relevant relativistic effects, as well as a
well optimized configuration interaction expansion, both of which
are important to obtain precise A-values for fine structure transitions.

\begin{figure*}
\epsscale{2.0}
\plottwo{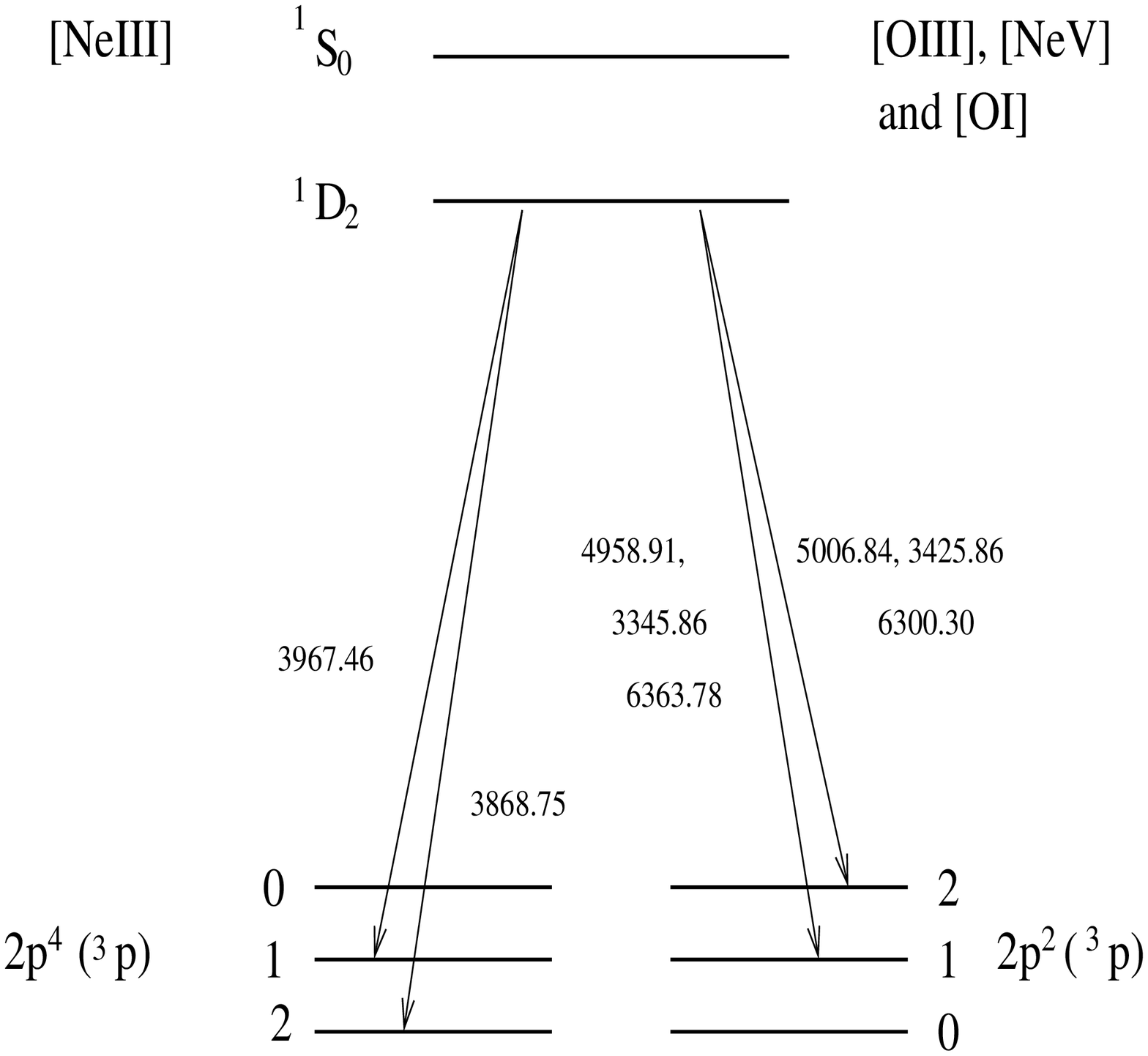}{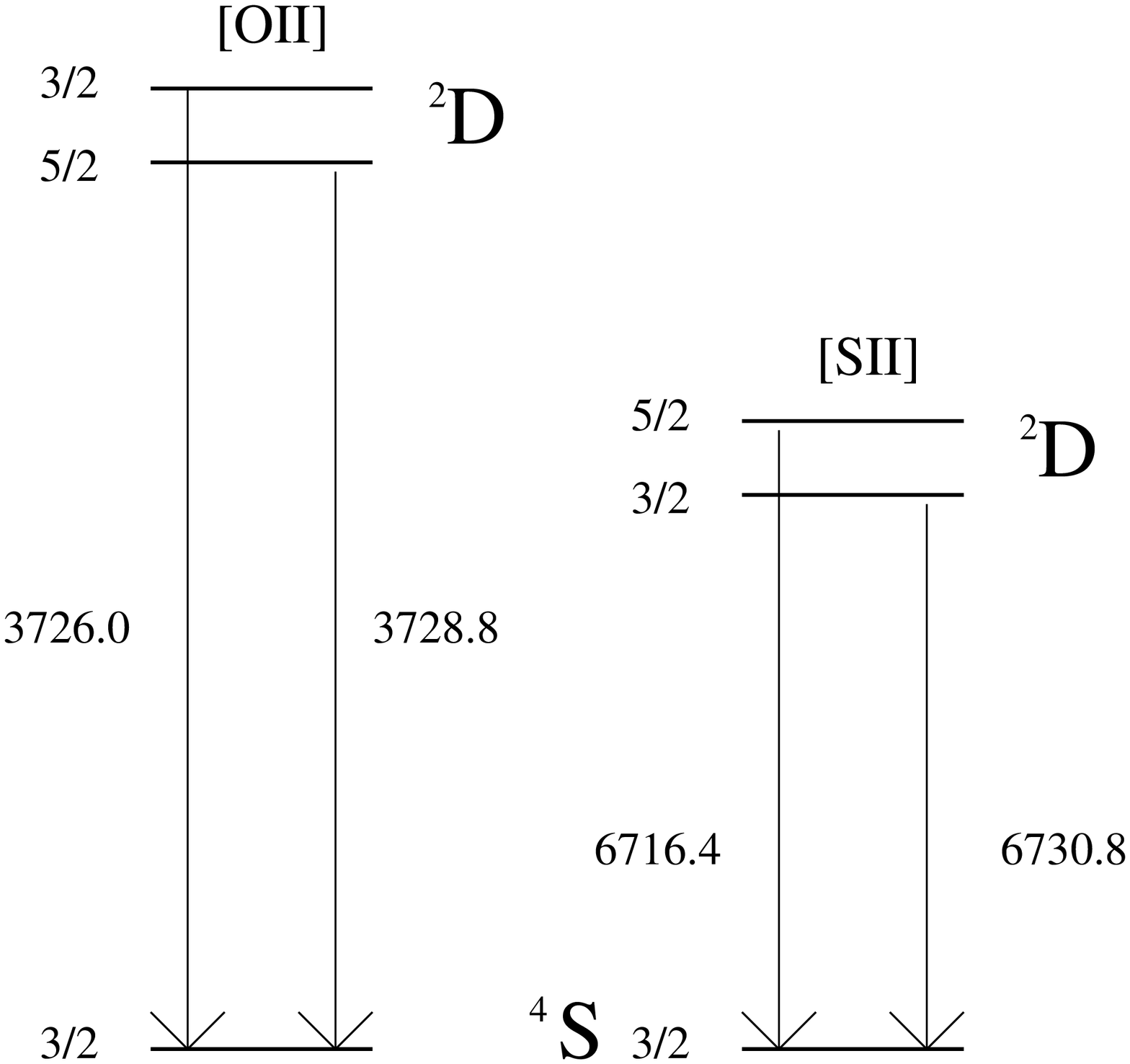}
\caption{\label{linescheme} Schematic diagram of the [OIII], [NeIII], and [NeV]
(left panel), and the [OII] and [SII] (right panel)
line doublets
}
\end{figure*}

 The actual value of the forbidden line ratio may be of decisive
importance in spectral identification and analysis of observational
datasets, and they indicate the extent of line blending and
signal-to-noise ratio. It is therefore instructive to examine further
(with a view towards future work as well) the line ratios of interest in
this work. For the forbidden lines in the left panel of
Figure\,\ref{linescheme}
the dominant contribution
is from the M1 operator (the E2 contribution is about 3 orders of
magnitude smaller). In the LS coupling limit, as the magnetic
interaction goes to zero, the ratio of the line strengths
$S(^1D_2-^3P_2)/S(^1D_2-^3P_1)$ = 3 \citep{sz00}. The A-values
however involves the energy differences as well, i.e.
$A(^1D_2-^3P_2)/A(^1D_2-^3P_1)$ = 3 $E(^1D_2-^3P_2)3/E(^1D_2-^3P_1)3$.
The line ratio (Eq. 3) deviates from the value of 3, in accordance
with the magnitude of the magnetic interaction. It is highly fortuitous
that for O~III this ratio is in fact 3, as noted by \citet{bahcall04} and 
\citep{sz00}.
But for other ions there is significant deviation, as seen in
Table\,\ref{line_ratio}. We do not give line ratios for the [SII] lines because
their line ratios depends strongly in the gas temperature 
\citep[e.g.][]{oster89}.
We derive this ratio for all three ions using 4 different datasets
available in literature. While all sets of line ratios agree to within
few percent, we envisage that this level of difference could turn out to
be crucial in analyzing large datasets, with line ratios
measured and calculated to better than 1\% accuracy.

 In addition to the ions with ground state fine structure levels
discussed above, there are well known pairs of lines in ions with excited state
fine structure levels, such as the doublet forbidden lines [O~II] and [S~II]
at $\lambda\lambda(\AA)$ = (3728.8/3726.0) and (6716.5/6730.8)
respectively (right panel of Figure\,\ref{linescheme}),
observed from many H~II regions and AGN \citep{oster89, p76}.
While collisional coupling
leads to variations in line ratios dependent on electron density,
the limiting values of these ratios are precisely known.
For example, the low density
line ratio LR (N$_e \longrightarrow 0$) = 1.5 for both [O~II] and [S~II]:
the ratio of the statistical weights of the upper levels in associated
pair of transitions ($^2D^o_{5/2},^2D^o_{3/2} \longrightarrow ^4S^o_{3/2})$.
The high density limit is given by

\be
    LR (N_e \rightarrow \infty) = \frac{g(^2D^o_{5/2})}{g(^2D^o_{3/2})}
\frac{A((^2D^o_{5/2}-^4S^o_{3/2})}{A(^2D^o_{3/2}-^4S^o_{3/2})},
\ee

and is = 0.35 for [O~II] \citep{z87} and 0.44 for [S~II]
\citep{mz82}.
Studies of ionized gaseous nebulae show no known cases of deviations from
these 'canonical' limits \citep{wang04}.
Whereas the wavelength separation
between the [O~II] doublet is possibly too small to be resolved in AGN,
the [S~II] doublet is easily observed and resolved, as in the present
exploratory work.

\section{\label{observe} Observations and data reduction}

We observed a small sample of 14 Seyfert 1.5 galaxies with the 2.4m Hiltner
telescope at MDM observatory at Kitt Peak, Arizona, 
for 5 nights from 2003-10-16 to 2003-10-21 to cover the [NeV],
[NeIII] and [OIII] lines --- the blue part of the optical spectrum --- 
and 3 nights starting 2004-10-10 to observe the
[OI] and [SII] doublets towards the red part of the spectrum.
(Table\,\ref{obs_log}). All spectra were taken with the OSU CCDS spectrograph
with the 600 grooves/mm grating in first order.
The slit width of the 2003 run was 1$^{''}$
corresponding to a spectral resolution in FWHM of about 2\AA. The weather
conditions during the whole run were excellent with mostly photometric
conditions. The slit position was normally at E-W for the spectra in the
4000-7000\AA~ range, but all spectra in the blue were observed in N-S direction
to compensate for refraction losses in the earth's atmosphere.
During the 2004 run the weather conditions were clear but suffered from rather
bad seeing. Most observations during the 2004 run were performed with  slit
widths of 1.5$^{''}$ and 2.0$^{''}$. The slit was oriented in N-S direction  for
all 2004 observations.

The total exposure times per spectrum are given in Table\,\ref{obs_log}. Each
spectrum consists of 2-4 single spectra which were combined after going through
all data reduction steps. For each individual spectrum a wavelength calibration
spectrum was taken.  
For the wavelength calibration we used a Hg lamp for the 3400-4300\AA~range, Xe
for the 4100-5100\AA~range, Ne for the 5100-6000\AA~ range, and
Ar for the 4900-5900\AA~and all observation with $\lambda>$6000\AA.
For the flat field correction the
CCDS spectrograph only allows  internal flats. To perform a flat field
correction we used an average of 10 flats.
 We used standard stars
BD+28-4211, Feige 110, G191 B2B, and Hiltner 600 for the flux calibration.
The data reduction was performed with the ESO MIDAS data reduction and analysis
package version 01FEBpl1.4.

The wavelength measurements were performed by fitting a Gaussian at 80\% of the
line peak  in order to avoid possible line asymmetries that
can occur at lower parts of an emission line. Line fluxes were measured by
integrating over the
whole line using the MIDAS command {\it integrate/line}. In order to estimate
the error of the data reduction, the data were reduced and analyzed
independently by D.G. and S.F..

\section{\label{results} Results}

\begin{figure*}
\caption{\label{opt_spectra} Optical spectra of the [NeV], [NeIII], and [OIII]
regions
}
\epsscale{2.6}
\plotone{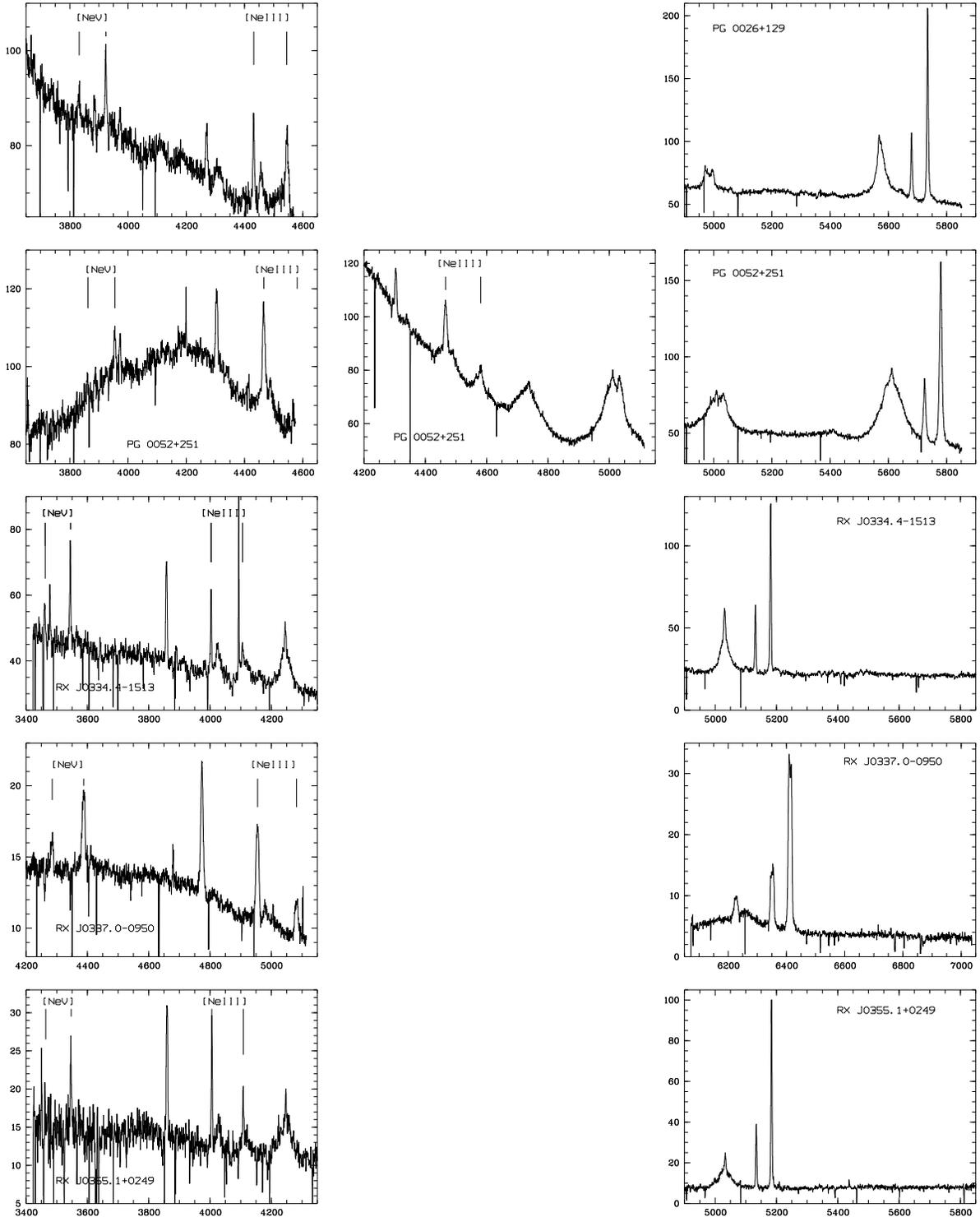}
\end{figure*}

\begin{figure*}
\caption{\label{opt_spectra_sii} Optical spectra of the [OI] and [SII] line
regions
}
\epsscale{2.6}
\plotone{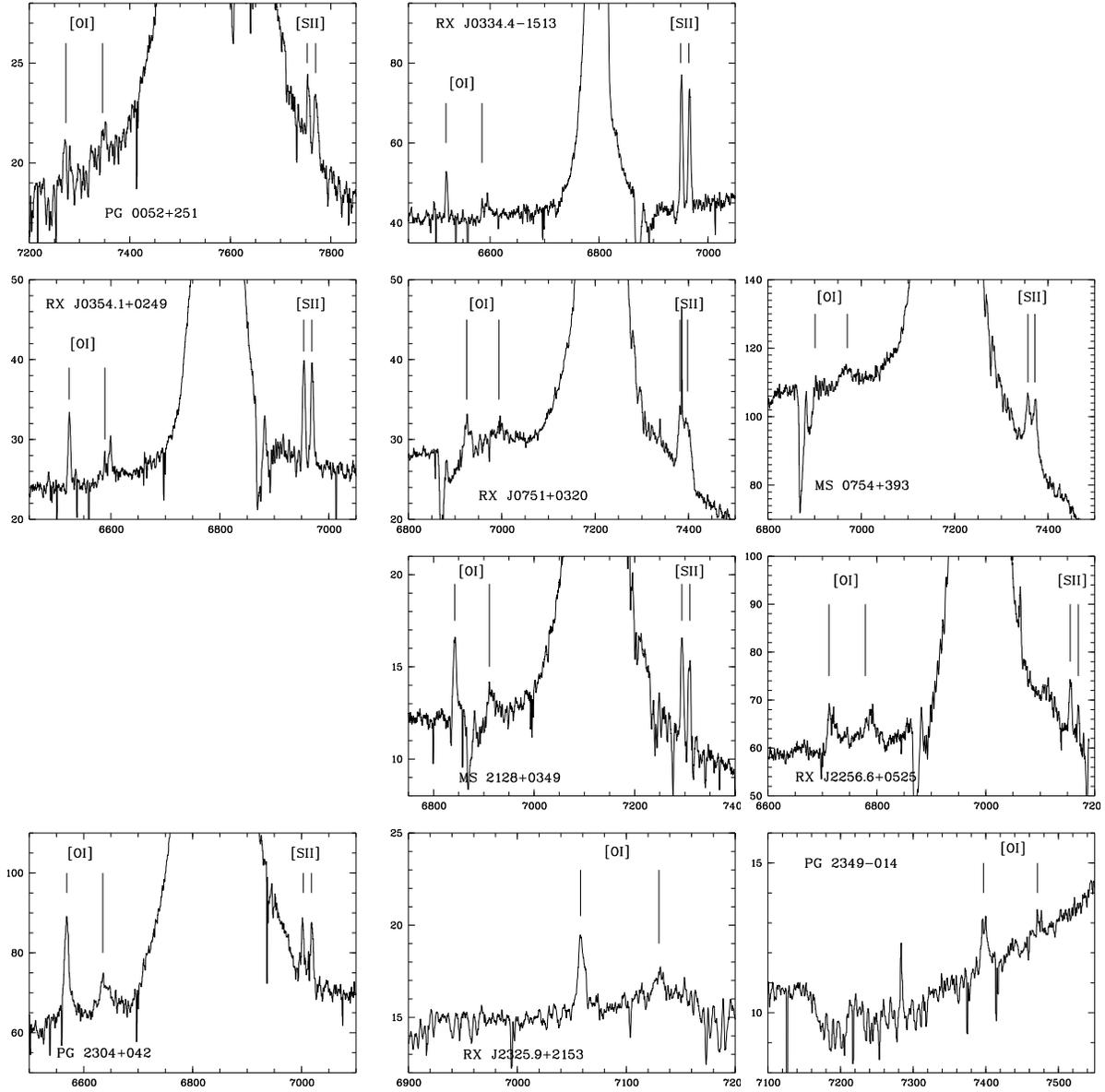}
\end{figure*}

Figure\,\ref{opt_spectra} displays the optical spectra of the [NeV0, {NeIII],
and [OIII] line regions of the objects sorted by RA as given in
Table.\,\ref{obs_log}.  As shown in
Figure\,\ref{opt_spectra} in most of the sources the lines are clearly
present and have enough S/N that their wavelengths and fluxes
can be measured. The spectra of the [OI] and [SII] lines taken in October 2004
are shown in Figure\,\ref{opt_spectra_sii}. In most cases the [SII] lines are
clearly present. However, the [OI]$\lambda$6363 line in most cases is too noisy
to allow accurate measurements of the wavelength.

Table\,\ref{lineflux} lists the line flux of the [OIII]$\lambda$5007\AA~ line and the 
line
ratios. In general, the [NeV]$\lambda$3426 and [NeIII]$\lambda$3869 line 
are about 1/10th of
the [OIII]$\lambda$5007 line.
Table\,\ref{linewave} lists the observed wavelengths of the [NeV], [NeIII],
[OIII], [OI], and [SII] lines.

For the low-redshift AGN in our sample with a maximum look-back time of about 2
Gyrs we would expect a maximum change of $\alpha$ of 4$\times 10^{-6}$ regarding
the the laboratory measurements of \citet{peik04}. Therefore the expected ratio
$\alpha^2(t)/\alpha^2(t)$ maximum is 1.000008. Every deviation from this value
gives us a handle on the quality of our measurements. 
Table\,\ref{alpha_ratio} lists the ratios of $\alpha^2(t)/\alpha^2(0)$ 
which are equal  to
the R(t)/R(0) ratio,
 with R(t) = $(\lambda2(t)-\lambda1(t))/(\lambda1(t) + \lambda2(t)$ and
R(0) are given by the laboratory wavelengths as R(0)([NeV])=1.1814, 
R(0)([NeIII])=12.597, R(0)([OIII])=4.80967, R(0)([OI])=5.011971,
and R(0)([SII])=1.0690657 $\times 10^{-3}$ (for all lines).
Table\,\ref{alpha_ratio} also summarizes the mean values of 
$\alpha^2(t)/\alpha^2(0)$ for each object as well 
the sample averaged value. In general, we find results with the lowest error 
estimates from the [NeIII] lines owing to their wide wavelength separation, 
whereas the values obtained from the [SII] lines suffer in precision from 
their small splitting. Clearly, the lines with the highest SNR offer the best 
possibilities to constrain their centers, making the original approach based 
upon the [OIII] doublet so effective.}

\begin{figure*}
\epsscale{1.2}
\plotone{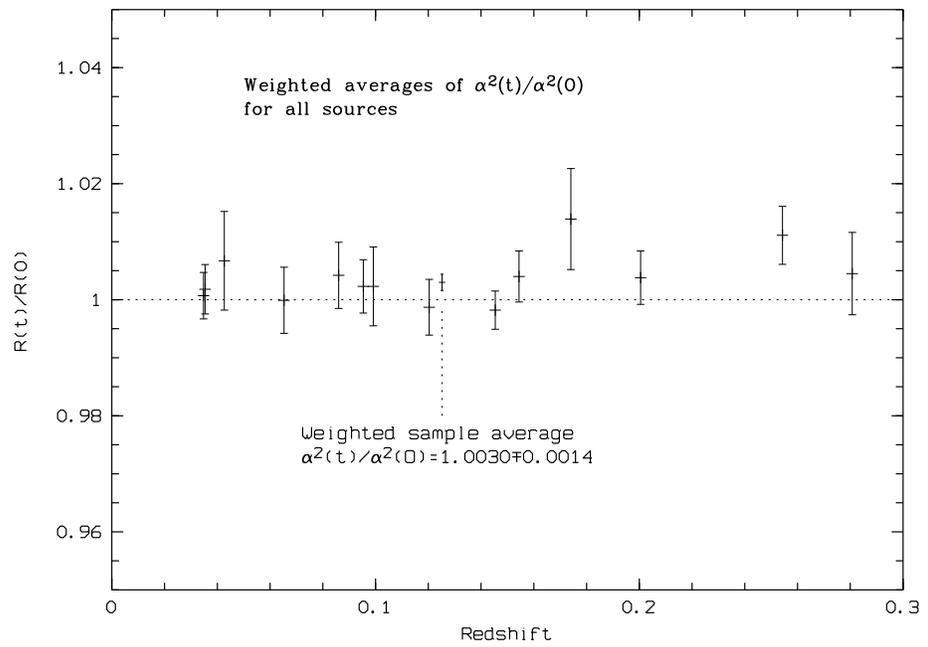}
\caption{\label{alpha-z} Redshift vs. $\alpha^2(t)/\alpha^2(0)$
}
\end{figure*}

Even though we are not attempting to measure a redshift dependence of the fine
structure constant $\alpha$ by redshift due to the low redshift of the AGN in
our sample, Figure\,\ref{alpha-z} displays the redshift vs.
$\alpha^2(t)/\alpha^2(0)$ diagram. As expected we do not see any significant
deviation from 1.0000.

\section{\label{discuss} Discussion}

As mentioned before, it is beyond the scope of this paper to measure the
time dependence
of the fine structure constant $\alpha$ with adequate precision. 
The redshifts of the objects in our sample
only cover a range between 0.034-0.281. The look-back time of an object of z=0.28 is of
the order of $\approx$ 3 Gyrs. With the upper limit of a possible change of
\alp~measured by \citet{peik04} of $\Delta\alpha = 2\times 10^{-15}$ yr$^{-1}$
we would expect an upper limit of $\Delta\alpha=6\times 10^{-6}$, which implies 
that \textbf{a}{} two orders of magnitude improvement is needed over what
we can achieve from our current data set.

For measuring any redshift dependence of the fine structure
constant \alp~,  estimates of the  accuracy in determining the wavelengths of the line
doublets are absolutely crucial. Evaluating $\alpha^2(t)/\alpha^2(0)$ 
we do not need to rely on absolute wavelength calibrations, but relative 
wavelengths ratios. However, the
issues that are important are: First, the stability of the wavelength 
calibration
 between
two lines in a line pair, which can be separated up to $\approx$100\AA~ in the 
case of
[NeIII], and second, the ability to determine the line centroids. While the
first issue is relatively safe for the [OIII], [OI] and [SII] lines, because their
separation is rather small and the wavelength calibration is very secure due to
the large number of calibration lamp lines, the situation is more difficult for
the [NeIII] and [NeV] lines. Not only that the separation between the two lines in
the [NeIII] and [NeV] line pairs are about 100\AA, also the wavelength calibration
in the blue using the CCDS spectrographs suffers from a lack of enough calibration
lamp lines. The CCDS spectrograph only has a Hg lamp providing just 5 calibration
lines throughout the blue/UV wavelength range. 
Certainly, a Th/Ar lamp could improve our ability to provide for a more 
stable calibration in this regime. The situation for these line pairs would 
also improve for objects of higher redshift as the observed features shift 
into the longer wavelength regime where the calibration is better.
 Furthermore, we indicate that the abundant night-sky lines upwards from 
 6500 \AA{} could be used as additional calibrators to measure small-scale
  flexures  as already indicated by \citep{bahcall04}.

 Measuring the center of the lines depends on a variety of factors: the spectral
resolution and dispersion of the instrument, the pixel sampling, the 
signal-to-noise ratio in the line and the line shape that often deviates 
substantially from a simple Gaussian. With the setup used for our observations,
 the resolution is
$\lambda/\Delta\lambda$=2000 and the dispersion yields 0.79 \AA{} per pixel. 
A low S/N also introduces an error in the wavelength
measurement, because noise changes the shape of the line and causes the 
estimate of the line
center to shift. Clearly, lines with high SNR will provide the best results
 regarding this aspect.\\
An important aspect of performing a high-precision analysis study is the 
sample selection.
\citet{bahcall04} had to be very critical about the shape of the [OIII]
lines of the sources in his sample reducing the number of good targets from
about 1000 to about 40. This was necessary for the SDSS sources which were
reduced and analyzed automatically. For our small sample, we could work on each 
source manually and had hoped to obtain reasonably good lines from prior 
knowledge. Our sources were selected to be X-ray hard which tend to have
 stronger [OIII] emission than
soft X-ray selected AGN \citep{gru04a}. For many of the sources optical spectra
were available from the ROSAT Bright Source Catalogue
(www.aip.de/$\sim$aschopwe/rbscat/rbscat.html){} which gave us some 
indication of the expected line shapes. Furthermore, by only fitting a 
Gaussian line to the narrow part of the emission lines, we tried to minimize 
the effect of asymmetric line shapes as much as possible.

 The identification and analysis of lines is greatly facilitated by the fact
 that the emission line multiplets chosen in our study have well constrained 
 line ratios. Therefore observed line pairs with ratios outside the 
 theoretical limits can be safely ruled out if they are blended or suffer 
 from instrumental effects. However, it is essential that the relevant 
 Einstein A-coefficients be accurately calculated, particularly taking account 
 of all relativistic effects. Whereas such data are available for the ions 
 and transitions considered in this work, high precision atomic calculations 
 are needed before expending the study to more complicated atomic species. 
 If these calculations are carried out then emission line studies of other 
 ions should be possible.\\  
Taking these factors into account individually for each line pair, we have 
estimated an error budget for each line centroid measurement which is listed 
in table\,\ref{linewave}. 
On average, we thus believe to be able to constrain an 
individual line center to about 0.6\AA, a value substantially higher than 
\citep{bahcall04} who estimate their precision to 0.05 pixels or about 
0.06 \AA{} at 5000 \AA. The effect on the error budget of 
${\alpha (t)}^2/{\alpha (0)}^2 \sim R(t)/R(0)$ depends on the actual
 wavelength splitting of the line pairs, and is most pronounced for the 
 lines with small separations. We achieve the best values for the [NeIII] 
 lines, but even in that case the error on an individual measurement of
  R(t)/R(0) is never below $1.0 \times 10 ^{-3}${} and usually of the order 
  $7 \times 10 ^{-3}${}.

For the time being the primary aim of this study is to lay the 
 groundwork for future
studies based on the many emission-line method with 8-10 m telescopes to
explore the high-z regime of faint objects. For example, the 
Large Binocular Telescope (LBT) is
slated to have two spectrographs, the Multi-Object Double Spectrograph
\citep[MODS\footnote{http://www.astronomy.ohio-state.edu/LBT/MODS/}][]{osmer00}
in the optical 0.3-1.0 $\mu$m range, and the other 
LUCIFER\footnote{http://www.lsw.uni-heidelberg.de/projects/Lucifer/index.html}
 in the J,H,K bands. Present observations were
divided into two parts, one focusing on ions in the blue side ([NeV],
[NeIII], and the other on
ions in the red side of the optical spectrum ([OI] and [SII]), with [OIII]
in the middle. As the higher-z objects become accessible with, say, the
LBT, pairs of lines from these ions would move into the range from
MODS into the J,H,K bands covered by LUCIFER. This would enable a
natural and logical extension of the present studies with LBT.\\
With the predicted capabilities of LUCIFER, MODS at the LBT, we estimate 
the error budget of an individual line measurement of [NeIII] at a redshift 
of z$\sim$2.5 to be $\sim$0.2 \AA~ which allows us to constrain 
${\alpha (t)}^2/{\alpha (0)}^2${} to $8 \times 10^{-4}${} even with our much
 simpler fitting approach than \citep{bahcall04}{} used for their sample. 
 Implementing their detailed analysis for the determination of the line 
 centroid and carefully choosing a reasonably sized sample with prior 
 line-shape knowledge, we will be able to push the error limit for an 
 averaged ${\alpha (t)}^2/{\alpha (0)}^2${} to $\sim 10^{-5..6}${}, 
 comparable to the limit reached in recent absorption line studies and thus 
 providing an interesting alternative.

In summary, the emission line multi-multiplet method has certain advantages: it
combines the multi-multiplet absorption line method, having many line pairs and
getting good statistics of the measurements of  
$\alpha^2(t)/\alpha^2(0)$, with
the ease of identifying lines and measuring the wavelengths of the
emission line pairs. It is observationally relatively inexpensive
 to get good spectra of the [OIII] and [NeIII] line pairs. 
 However, our experience
 with the current data set shows that by using a 2m class telescope in order to
 archive better S/N of the [NeIII] and [NeV] lines more than the typical
  1-2 hours
 of observing time that we spent have to be invested. Is this project still
 suitable or 2-3m class telescopes? In principle, yes. 
 However it becomes more and more
 challenging for objects with higher redshifts which are fainter and therefore
 require much longer integration times. Because a possible time dependence of the
 fine structure constant \alp~can only be measured from quasars with
 redshifts of z=3 or higher this type of high-precision measurements is limited
 to large telescopes with medium- or high-resolution
 NIR spectrographs only. However, because the resolution does not 
 have to be as
 high as for the absorption line method, the exposure time per object will be in
 the order of an hour only. 
 Nevertheless smaller telescopes
 can build up the fundamentals which can be extended by larger
 telescopes. With a number of 8-10 m class telescopes available in the
 future it should be feasible to request allocation for such a project.

\section{\label{conclude} Conclusion}

We have demonstrated the application of a method for 
studying time variation of the fine structure constant based on the analysis 
of many emission line multiplets, with the following salient features:

\bl~ Extension of the [O III] multiplet method by 
\cite{bahcall04} to a `many multiplet method' of 
well known forbidden line multiplets in the optical 
rest frame. 

\bl~ Our best measurements archive errors in $\alpha^2(t)/\alpha^2(0)$ in the
order of $10^{-3}$.

\bl~ In contrast to \citet{bahcall04} whose work based on a search of the SDSS
 database, the present work involved new observations of selected Seyfert 1 
 galaxies 
specifically targeted to obtain several emission line multiplets. Two 
separate observation runs were made, focused on the blue and the red sides 
of the optical spectrum.

\bl~ Best suited are sources with strong very narrow NLR lines. However, this
requirement becomes more challenging for high-redshift AGN. High redshift AGN
tend to have central black holes with masses in the order of $10^9$ to
$10^{10}$\msun \citep[e.g.,][]{die04, vester04}. Due to the well-known relation
between the black hole mass and the bulge stellar velocity distribution, the
$M_{\rm BH} - \sigma$ relation \citep[e.g.,][]{ferr00, geb00}, high-redshift
quasars will tend to have broader NLR emission lines than low-redshift AGN which
tend to have smaller black hole masses. 
Counteracting this problem is the growing wavelength separation for increasing
 redshifts.

\bl~ If a Thorium calibration lamp is not available to observe at wavelength
$\lambda<$4100\AA, it is recommended to use only AGN with a redshift z$>$0.24 to
observe the [NeV] lines with appropriate wavelength calibration.

\bl~ Analysis of observed pairs of lines showed the necessity of 
high accuracy theoretical atomic calculations in order to obtain line 
ratios which, in principle, can be ascertained a priori.

\bl~ The next generation of 8-10m class of telescopes should be able 
to achieve the required precision of $\Delta \alpha(t) \sim $ 10$^{-5..-6}$ 
to make a more definitive prediction.

\acknowledgments

We would like to thank Axel Schwope (AIP) for making his ROSAT Bright Source
Catalog available to us. We would also like to thank Bob Barr and his crew
at MDM observatory for their technical support.
This work was supported in part by a grant from the astronomy division of the
National Science Foundation.



\begin{deluxetable}{lll}
\tablecaption{Rest-frame wavelength and atomic transitions
of the line doublets \label{line_wave}}
\tablehead{
\colhead{Line} & \colhead{$\lambda$ [\AA]} & \colhead{Atomic transition}
}
\startdata
$\rm [Ne V]$   &  3345.86   & $\rm ^1D_2 - ^3P_1$ \\
$\rm [Ne V]$   &  3425.86   & $\rm ^1D_2 - ^3P_2$ \\
$\rm [Ne III]$ &  3868.75   & $\rm ^1D_2 - ^3P_2$ \\
$\rm [Ne III]$ &  3967.46   & $\rm ^1D_2 - ^3P_1$ \\
$\rm [O III]$   &  4958.91   & $\rm ^1D_2 - ^3P_1$ \\
$\rm [O III]$   &  5006.84   & $\rm ^1D_2 - ^3P_2$ \\
$\rm [O I]$     &  6300.304  & $\rm ^1D_2 - ^3P_2$ \\
$\rm [O I]$     &  6363.776  & $\rm ^1D_2 - ^3P_1$ \\
$\rm [S II]$   &  6716.440  & $\rm ^2D^o_{5/2} - ^4S^o_{3/2}$ \\
$\rm [S II]$   &  6730.816  & $\rm ^2D^o_{3/2} - ^4S^o_{3/2}$
\enddata

\end{deluxetable}

\begin{deluxetable}{cccccc}
\tablecaption{Line Ratios For [O III], [Ne V] And [Ne III] Using
Different Atomic Data. \label{line_ratio}}
\tablewidth{0pt}
\tablehead{
\colhead{Ion} &
\colhead{$\Delta E(^1D_2-^3P_2)$} &
\colhead{$\Delta E(^1D_2-^3P_1) $} &
\colhead{$ A(^1D_2-^3P_2) $} &
\colhead{$ A(^1D_2-^3P_1) $} &
\colhead{LR}
}
\startdata
 O III & 0.18195 & 0.18371 & 1.8105(-2) & 6.212 (-3) & 2.89$^a$ \\
       &  "      & "       & 1.96(-2)   & 6.74  (-3) & 2.88$^b$ \\
       &  "      & "       & 2.041(-2)  & 6.995 (-3) & 2.89$^c$ \\
       &  "      & "       & 2.042(-2)  & 6.785 (-3) & 2.98$^d$ \\
 Ne V  & 0.26592 & 0.27228 & 3.82(-1)   & 1.38(-1)   & 2.70$^a$ \\
       & "       & "       & 3.65(-1)   & 1.31(-1)   & 2.72$^b$ \\
       & "       & "       & 3.499(-1)  & 1.252(-1)  & 2.73$^c$ \\
       & "       & "       & 3.501(-1)  & 1.221(-1)  & 2.80$^d$ \\
 Ne III & 0.23548 & 0.22962 & 1.703(-1) & 5.24(-2) & 3.33$^a$ \\
        & "       & "       & 1.71(-1)  & 5.42(-2) & 3.24$^b$ \\
        & "       & "       & 1.73(-1)  & 5.344(-2) & 3.32$^c$ \\
        & "       & "       & 1.708(-1)  & 5.413(-2) & 3.24$^d$
\enddata

Notes: a - NIST Compilation, b - Pradhan and Peng Compilation (1995), c
- Galavis et al. (1997), d - Storey and Zeippen (2000).

\end{deluxetable}

\begin{deluxetable}{rlccllccc}
\tabletypesize{\scriptsize}
\tablecaption{Observation log, observing times are given in minutes
\label{obs_log}}
\rotate
\tablewidth{0pt}
\tablehead{
& & & & & & \multicolumn{3}{c}{$\rm T_{obs}$} \\
\colhead{\rb{\#}} &
\colhead{\rb{Object}} & \colhead{\rb{RA-2000}} &
\colhead{\rb{DEC-2000}} &
\colhead{\rb{B-mag}} & \colhead{\rb{z}} & \colhead{[NeV], [NeIII]} &
\colhead{[OIII]} & \colhead{[OI], [SII]}
}
\startdata
1 & PG 0026+129 & 00 29 14 & +13 16 04 & 15.41 & 0.14537 &
60 & 30 & --- \\
2 & PG 0052+251 & 00 54 52 & +25 25 39 & 15.42 & 0.15439 & 60, 80 & 30 & 60\\
3 & RX J0334.4$-$1513 & 03 34 24 & $-$15 13 40 & 15.43 & 0.03478 &
90 & 40 & 120 \\
4 & RX J0337.0$-$0950 & 03 37 03 & $-$09 50 02 & 17.0 & 0.28074 &
110 & 60 & 120 \\
5 & RX J0354.1+0249 & 03 54 09 & +02 49 30 & 16.3 & 0.03536 &
80 & 60 & 120 \\
6 & RX J0751.0+0320 & 07 51 00 & +03 20 41 & 15.2 & 0.09914 &
30 & 25 & 90 \\
7 & MS 0754+393 & 07 58 00 & +39 20 49 & 14.36 & 0.09533 & 60, 60 & 20
& 120 + 40 \\
8 & RX J0836.9+4426 & 08 36 59 & +44 26 02 & 15.6 & 0.25427 &
90 & 30 & 75 \\
9 & MS 2128.3+0349 & 21 30 53 & +04 02 30 & 16.34 & 0.08600 &
120 & 60 & 120 \\
10 & PKS 2135$-$147 & 21 37 45 & $-$14 32 55 & 15.91 & 0.20048 & 80, 60 & 40
& 90 \\
11 & RX J2256.6+0525 & 22 56 37 & +05 25 16 & 16.2 & 0.06529 &
120 & 40 & 60 \\
12 & PG 2304+042 & 23 07 03 & +04 32 57 & 15.44 & 0.04265 &
60 & 45 & 80 \\
13 & RX J2325.9+2153 & 23 25 54 & +21 53 16 & 15.9 & 0.12033 &
80 & 60 & 120 \\
14 & PG 2349$-$014 & 23 51 56 & $-$01 09 13 & 15.7 & 0.17404 & 60, 80 & 50 & 75
\enddata

\end{deluxetable}

\begin{deluxetable}{rlrlllllllll}
\tabletypesize{\scriptsize}
\tablecaption{Line fluxes (observer's frame).
\label{lineflux}}
\tablewidth{0pt}
\rotate
\tablehead{
\colhead{\#} &
\colhead{Object} & 
\colhead{$F_{\rm [OIII]4959}$\tablenotemark{1}} & 
\colhead{$F_{\rm [OIII]5007}$} &
\colhead{$F_{\rm [NeV]3346}$} &
\colhead{$F_{\rm [NeV]3426}$} &
\colhead{$F_{\rm [NeIII]3869}$} &
\colhead{$F_{\rm [NeIII]3967}$} &
\colhead{$F_{\rm [OI]6300}$} &
\colhead{$F_{\rm [OI]6364}$} &
\colhead{$F_{\rm [SII]6716}$} &
\colhead{$F_{\rm [SII]6730}$} 
}
\startdata
1 & PG 0026 & 304$\pm$30 & 1014$\pm$40 & 20$\pm$ 10 & 83$\pm$ 20 & 90$\pm$ 20 & 30$\pm$ 10 & --- & --- & --- & --- \\
2 & PG 0052 & 320$\pm$ 30 & 1290$\pm$ 60 & 10$\pm$ 5 & 44$\pm$ 10 & 190$\pm$ 30 & 45$\pm$ 20 & --- & --- & 16$\pm$ 8 & 25$\pm$ 10\\
3 & RXJ0334 & 174$\pm$ 30 & 490$\pm$ 30 & --- & --- & 80$\pm$ 5 & 20$\pm$ 10 & 68$\pm$ 8 & 20$\pm$ 10 & 90$\pm$ 15 & 20$\pm$ 5\\
4 & RXJ0337 & 126$\pm$ 20 & 400$\pm$ 20 & 10$\pm$ 5 & 50$\pm$ 10 & 64$\pm$ 8 & 28$\pm$ 5 & 45$\pm$ 10 & --- & --- & ---\\
5 & RXJ0354 & 130$\pm$ 6 & 413$\pm$ 15 & 6$\pm$ 4 & 26$\pm$ 10 & 55$\pm$ 10 & 18$\pm$ 4 & 8.7$\pm$ 2.0 & 0.75$\pm$ 0.7 & 10.9$\pm$ 1.0 & 8.4$\pm$ 1.2\\
6 & RXJ0751 &  540$\pm$ 20 & 1800$\pm$ 100 & 16$\pm$ 5 & 80$\pm$ 10 & 120 $\pm$20 & 50 $\pm$30 & 7 $\pm$4 & --- & --- & ---\\
7 & MS 0754 & 980$\pm$ 40 & 3790$\pm$ 100 & 34$\pm$ 10 & 250$\pm$ 50 & 380$\pm$ 30 & 500$\pm$ 30 & --- & --- & 110$\pm$ 30 & 96$\pm$ 4\\
8 & RXJ0836 &  390$\pm$ 50 & 1200$\pm$ 60 & 54$\pm$ 10 & 125$\pm$ 10 & 240$\pm$ 20 & 110$\pm$ 20 & 25$\pm$ 10 & 2$\pm$ 2 & --- & ---\\ 
9 & MS 2128 & 140$\pm$ 15 & 480$\pm$ 20 & 10$\pm$ 5 & 40$\pm$ 10 & 50$\pm$ 10 & 24$\pm$ 4 & 21$\pm$ 12 & 1.5$\pm$ 1.0 & --- & --- \\
10 & PKS2135 & 360$\pm$ 10 & 1190$\pm$ 40 & 9$\pm$ 3 & 50$\pm$ 7 & 55$\pm$ 10 & 20$\pm$ 7 & --- & --- & --- & ---\\
11 & RXJ2256 & 250$\pm$ 10 & 670$\pm$ 30 & --- & 70$\pm$ 10 & 117$\pm$ 20 & 75$\pm$ 50 & 3.5$\pm$ 1.5 & --- & 9$\pm$ 2 & 5$\pm$ 3\\ 
12 & PG 2304 & 175$\pm$ 25 & 640$\pm$ 30 & --- & 15$\pm$ 10 & 53$\pm$ 10 & --- & 26$\pm$ 6 & 1.5 $\pm$1.0 & 7.5$\pm$ 3.0 & 11.2$\pm$ 4.0\\
13 & RXJ2325 & 146$\pm$ 15 & 520$\pm$ 20 & --- & 24$\pm$ 8 & 42.5 $\pm$5.0 & 19$\pm$ 7 & 25$\pm$ 5 & --- & --- & ---\\
14 & PG 2349 & 60$\pm$ 20 & 207$\pm$ 30 & --- & --- & 17.5$\pm$ 7.0 & 3.5$\pm$ 2.5 & 10.5$\pm$ 3.0 & 2.6$\pm$ 1.5 & --- & --- \\
\enddata

\tablenotetext{1}{In units of $\rm
10^{-19}~W~m^{-2}$.}

\end{deluxetable}

\begin{deluxetable}{rlcccccccccc}
\tabletypesize{\scriptsize}
\tablecaption{Observed wavelengths of the [NeV], 
[NeIII], and [OIII] lines in units of \AA
\label{linewave}}
\rotate
\tablewidth{23cm}
\tablehead{
\colhead{\#} &
\colhead{Object} &
\colhead{$\rm \lambda[NeV]3346$} &
\colhead{$\rm \lambda[NeV]3426$} &
\colhead{$\rm \lambda[NeIII]3869$} &
\colhead{$\rm \lambda[NeIII]3968$} &
\colhead{$\rm \lambda[OIII]4959$} &
\colhead{$\rm \lambda[OIII]5007$} &
\colhead{$\rm \lambda[OI]6300$} &
\colhead{$\rm \lambda[OI]6363$} &
\colhead{$\rm \lambda[SII]6716$} &
\colhead{$\rm \lambda[SII]6734$}
}
\startdata
1 & PG 0026 &  3833.8$\pm$0.4& 3923.5$\pm$0.4& 4431.4$\pm$0.4 & 4545.6$\pm0.3$ & 5679.7$\pm$0.3& 5734.7$\pm$0.3 & --- & --- & --- & --- \\
2 & PG 0052 &  3860.4$\pm$0.6 & 3954.7$\pm$0.5 & 4465.9$\pm$0.4 & 4579.9$\pm$0.5 & 5724.6$\pm$0.3 & 5779.9$\pm$0.1 & 7269.5$\pm$2.0 & 7350.4$\pm$1.5 & 7753.7$\pm$0.5 & 7768.6$\pm$0.5 \\
3 & RXJ0334 &  3460.9$\pm$0.4 & 3544.1$\pm$0.3 & 4003.5$\pm$0.4 & 4105.9$\pm$0.6 & 5131.4$\pm$0.3 & 5180.9$\pm$0.3 & 6517.9$\pm$0.4 & 6583.4$\pm$0.5 & 6949.3$\pm$0.4 & 6964.0$\pm$0.4 \\
4 & RXJ0337 &  4284.7$\pm$0.5 & 4387.5$\pm$1.0 & 4954.6$\pm$0.4 & 5082.0$\pm$1.0 & 6351.3$\pm$0.8 & 6412.5$\pm$1.5 & 8071.7$\pm$1.0 & --- & --- & --- \\
5 & RXJ0354 &  3461.7$\pm$0.6 & 3545.9$\pm$0.6 & 4005.8$\pm$0.5 & 4108.1$\pm$0.5 & 5134.2$\pm$0.2 & 5183.9$\pm$0.3 & 6522.1$\pm$0.4 & 6586.9$\pm$0.4 & 6952.7$\pm$0.0.5 & 6967.7$\pm$0.5 \\
6 & RXJ0751 &  3673.2$\pm$1.0 & 3762.8$\pm$1.0 & 4252.0$\pm$0.5 & 4361.1$\pm$0.4 & 5450.3$\pm$0.6 & 5503.2$\pm$0.7 & 6925.6$\pm$0.5 & 6993.8$\pm$2.0 & 7382.0$\pm$2.0 & 7395.1$\pm$2.0 \\
7 & MS 0754 &  3665.7$\pm$0.4 & 3753.2$\pm$0.3 & 4236.8$\pm$0.5 & 4345.6$\pm$0.5 & 5431.6$\pm$0.5 & 5484.2$\pm$0.3 & --- & --- & 7356.9$\pm$0.7 & 7372.8$\pm$0.8 \\
8 & RXJ0836 &  4194.8$\pm$0.6 & 4295.9$\pm$0.5 & 4852.1$\pm$0.5 & 4976.8$\pm$0.5 & 6219.8$\pm$0.4 & 6279.9$\pm$0.5 & 7904.8$\pm$1.5 & 7989.2$\pm$1.5 & 8424.8$\pm$1.0 & 8443.5$\pm$0.8 \\
9 & MS 2128 &  3631.8$\pm$1.0 & 3720.6$\pm$0.7 & 4201.4$\pm$0.6 & 4309.4$\pm$0.9 & 5385.4$\pm$0.4 & 5437.5$\pm$0.5 & 6842.4$\pm$0.4 & 6911.3$\pm$0.4 & 7294.2$\pm$0.5 & 7309.1$\pm$0.8 \\
10 & PKS2135 & 4015.9$\pm$1.0 & 4112.7$\pm$.3 & 4644.5$\pm$0.4 & 4763.4$\pm$0.5 & 5953.0$\pm$0.4 & 6010.6$\pm$0.4 & 7363.6$\pm$1.0 & --- & --- & --- \\
11 & RXJ2256 & 3563.7$\pm$1.0 & 3649.2$\pm$0.4 & 4121.6$\pm$0.4 & 4227.3$\pm$0.5 & 5282.8$\pm$0.6 & 5333.8$\pm$0.5 & 6712.0$\pm$0.7 & 6779.9$\pm$1.0 & 7154.8$\pm$0.4 & 7169.8$\pm$0.4 \\
12 & PG 2304 & ---	& 3572.2$\pm$1.0 & 4033.9$\pm$0.5 & --- & 5170.5$\pm$0.3 & 5220.5$\pm$0.3 & 6567.5$\pm$0.6 & 6633.8$\pm$0.7 & 7001.0$\pm$0.9 & 7017.3$\pm$0.8 \\
13 & RXJ2325 & 3748.6$\pm$0.7 & 3837.1$\pm$0.4 & 4334.2$\pm$0.4 & 4445.8$\pm$0.5 & 5555.5$\pm$0.4 & 5609.3$\pm$0.4 & 7058.7$\pm$0.7 & 7130.9$\pm$2.0 & --- & --- \\
14 & PG 2349 & ---	& ---	   & 4541.8$\pm$1.3 &  4660.5$\pm$1.1 & 5822.3$\pm$0.4 & 5878.4$\pm$0.3 & 7394.6$\pm$1.5 & 7470.1$\pm$1.5 & 7884.9$\pm$0.8 & 7903.5$\pm$0.8 \\

\enddata

\end{deluxetable}

\begin{deluxetable}{rlccccccc}
\tabletypesize{\scriptsize}
\tablecaption{Redshift-ordered Ratios of $\alpha^2(t)/\alpha^2(0)$ measured
from the [NeV], [NeIII], [OIII], [OI], and [SII] lines, and the 
weighted average of all line pairs.
\label{alpha_ratio}}
\rotate
\tablewidth{0pt}
\tablehead{
\colhead{\#} &
\colhead{Object} & z &
\colhead{$\frac{\alpha^2(t)}{\alpha^2(0)}$([NeV])} &
\colhead{$\frac{\alpha^2(t)}{\alpha^2(0)}$([NeIII])} &
\colhead{$\frac{\alpha^2(t)}{\alpha^2(0)}$([OIII])}  &
\colhead{$\frac{\alpha^2(t)}{\alpha^2(0)}$([OI])}  &
\colhead{$\frac{\alpha^2(t)}{\alpha^2(0)}$([SII])}  &
\colhead{Average $\frac{\alpha^2(t)}{\alpha^2(0)}$}
}
\startdata
3  & RXJ0334 & 0.03478 &  1.0054 $\pm$ 0.0069 &  1.0024 $\pm$  0.0071 & 0.9980  $\pm$ 0.0086 & 0.9975 $\pm$  0.0097 & 0.9883 $\pm$  0.0380 &  1.0007 $\pm$ 0.0040\\
5  & RXJ0354 & 0.03536 &  1.0171 $\pm$  0.0085 & 1.0009 $\pm$  0.0069 & 1.0015 $\pm$  0.0073 & 0.9863 $\pm$  0.0087  &  1.0079 $\pm$ 0.0475 & 1.0018 $\pm$ 0.0042 \\
12 & PG 2304 & 0.04265 &   ---    & --- & 1.0005 $\pm$  0.0085 & 1.0020  $\pm$ 0.0139 &  1.0876  $\pm$ 0.0755 & 1.0067 $\pm$ 0.0085  \\
11 & RXJ2256 & 0.06529 &  1.0034 $\pm$  0.0126 & 1.0051 $\pm$  0.0061 & 0.9988 $\pm$  0.0153 & 1.0041 $\pm$  0.0181 & 0.953 $\pm$  0.0369 & 0.9999 $\pm$ 0.0057 \\
9  & MS 2128 & 0.08600 &  1.0223 $\pm$  0.0141 & 1.0074 $\pm$  0.0101 & 1.0009 $\pm$  0.0123 & 1.0000 $\pm$  0.0082 & 0.9544 $\pm$  0.0604 & 1.0042 $\pm$ 0.0057 \\
7  & MS 0754 & 0.09533 &  0.9983 $\pm$  0.0057 & 1.0064 $\pm$  0.0065 & 1.0019  $\pm$ 0.0111 & ---     & 1.0097 $\pm$  0.0675    & 1.0023 $\pm$ 0.0046 \\
6  & RXJ0751 & 0.09914 &  1.0199 $\pm$  0.0161 & 1.0056 $\pm$  0.0059 & 1.0042 $\pm$  0.0175 & 0.9776 $\pm$  0.0302  & 0.8292 $\pm$  0.1790     & 1.0023 $\pm$ 0.0068 \\
13 & RXJ2325 & 0.12033 &  0.9875 $\pm$  0.0090 & 1.0009 $\pm$  0.0057 & 1.0019  $\pm$ 0.0105 & 1.0152 $\pm$  0.0298 & ---     & 0.9987 $\pm$ 0.0048 \\
1  & PG 0026 & 0.14537 &  0.9788 $\pm$  0.0062 & 1.0099 $\pm$  0.0044 & 1.0019  $\pm$ 0.0077 & ---    & ---     & 0.9982 $\pm$ 0.0033 \\
2  & PG 0052 & 0.15439 &   1.0214 $\pm$  0.0095 & 1.0005 $\pm$  0.0056 & 0.9945 $\pm$  0.0057 & 1.1041 $\pm$  0.0341 & 0.897 $\pm$  0.0426 & 1.0040 $\pm$ 0.0044 \\
14 & PG 2349 & 0.17404 &  ---     & 1.0240  $\pm$ 0.0147 & 0.9969 $\pm$  0.0089 & 1.0134 $\pm$  0.0285 & 1.1020 $\pm$  0.0713     & 1.0139 $\pm$ 0.0087 \\
10 & PKS2135 & 0.20048 &  1.0080 $\pm$  0.0109 & 1.0033 $\pm$  0.0054 & 1.0011 $\pm$  0.0983 & ---     & ---     & 1.0038 $\pm$ 0.0046 \\
8  & RXJ0836 & 0.25427 &  1.0079 $\pm$  0.0077  & 1.0072 $\pm$  0.0057 & 0.9997  $\pm$ 0.0107 & 1.0595 $\pm$  0.0266    & 1.0370 $\pm$  0.0710    & 1.0111 $\pm$ 0.0050 \\
4  & RXJ0337 & 0.28074 &  1.0034 $\pm$  0.0109 & 1.0077 $\pm$  0.0085 & 0.9970 $\pm$  0.0277 & ---     & ---     & 1.0045 $\pm$ 0.0071 \\ \\
   & \multicolumn{2}{l}{Sample average}             &   &  &  &  &  & 1.0030 $\pm$ 0.0014 \\
\enddata

\end{deluxetable}

\begin{figure*}
\epsscale{2.6}
\plotone{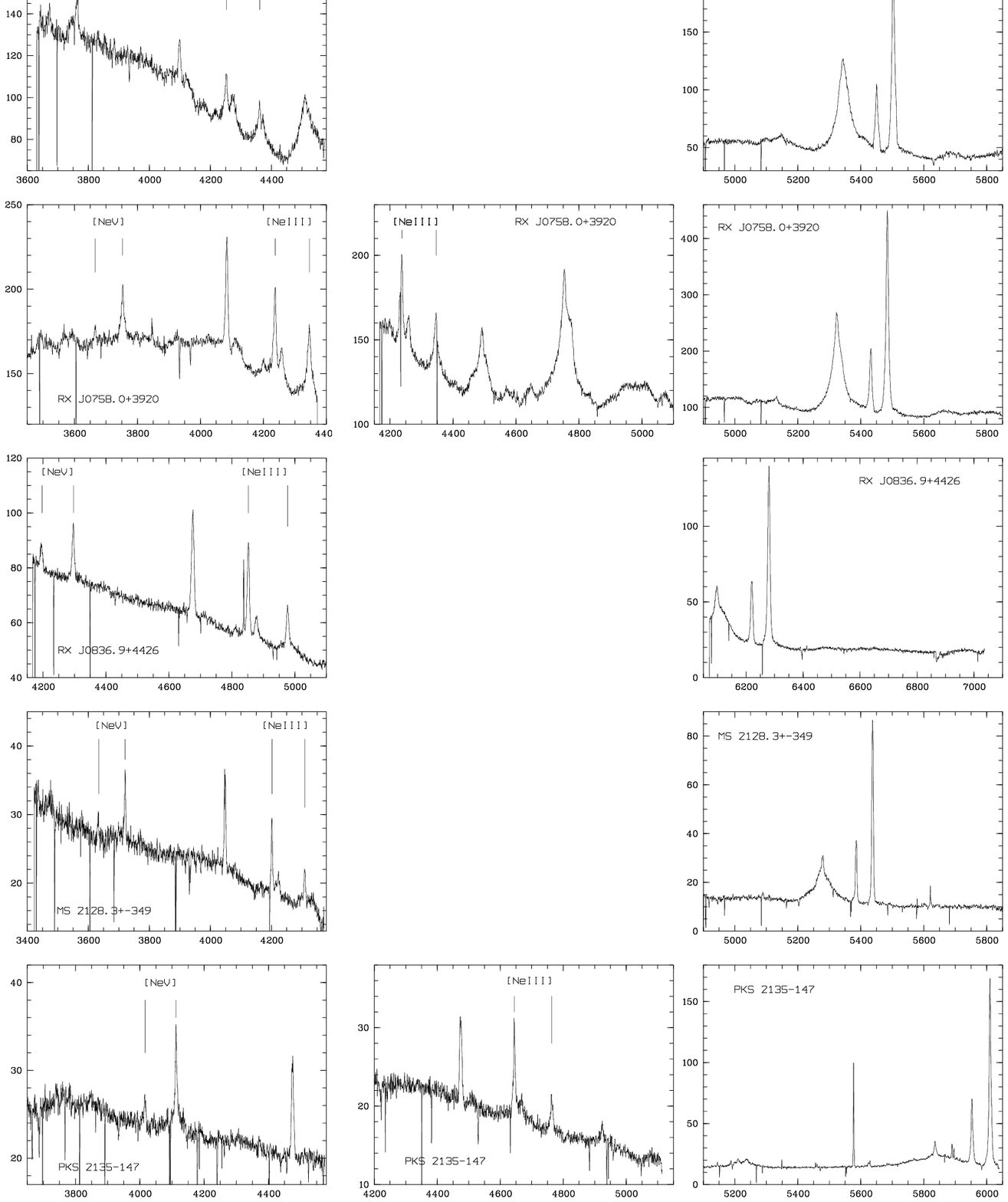}
\end{figure*}

\begin{figure*}
\epsscale{2.6}
\plotone{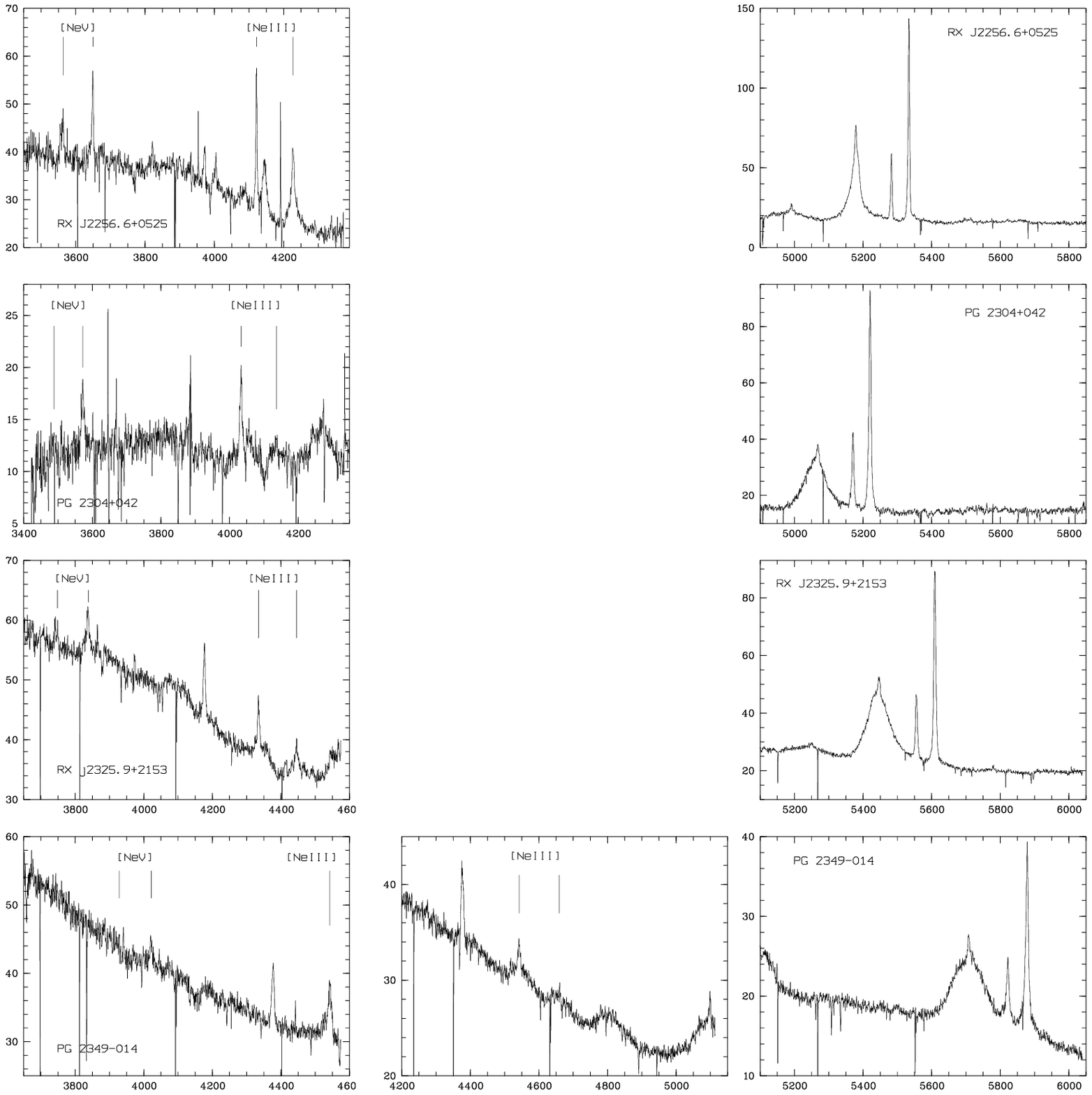}
\end{figure*}

\end{document}